\documentclass[a4paper,11pt]{article}
\usepackage{graphicx,amssymb,color,float}
\usepackage{epsfig}
\usepackage{color}
\usepackage{tabularx}
\usepackage[colorlinks=true,citecolor=blue, linkcolor=blue, urlcolor=blue]{hyperref}
\usepackage{amstext}
\usepackage{amssymb}
\usepackage{amsmath}
\usepackage{subfigure}
\usepackage{graphicx}
\usepackage{color}
\usepackage{ulem}
\usepackage[ruled, ]{algorithm2e}
\usepackage{booktabs}
\begin{document}

\centerline{\LARGE \bf
Estimation of time-varying recovery and death rates }

\medskip

\centerline{\LARGE \bf from epidemiological data: A new approach }

\medskip

\vspace*{1cm}

\centerline{\bf Samiran Ghosh$^{1}$, Malay Banerjee$^{2,3}$, Subhra Sankar Dhar$^{2,3}$, Siuli Mukhopadhyay$^{1,3,4}$}

\vspace{0.5cm}

\centerline{ $^1$  Department of Mathematics, Indian Institute of Technology Bombay, }
\centerline{  Mumbai 
- 400076, India}

\centerline{ $^2$  Department of Mathematics and Statistics, Indian Institute of Technology Kanpur,}

\centerline{  Kanpur - 208016, India}

\centerline{ $^3$  National Disease Modelling Consortium, Indian Institute of Technology Bombay, }

\centerline{  Mumbai 
- 400076, India}

\centerline{ $^4$ Koita Centre for Digital Health, Indian Institute of Technology Bombay,}

\centerline{  Mumbai 
- 400076, India}

\vspace{0.5cm}
\centerline{Corresponding author: Subhra Sankar Dhar (Email: subhra@iitk.ac.in)}

\vspace{2cm}

\noindent
 {\bf Abstract.} The time-to-recovery or time-to-death for various infectious diseases can vary significantly among individuals, influenced by several factors such as demographic differences, immune strength, medical history, age, pre-existing conditions, and infection severity. To capture these variations, time-since-infection dependent recovery and death rates offer a detailed description of the epidemic. However, obtaining individual-level data to estimate these rates is challenging, while aggregate epidemiological data (such as the number of new infections, number of active cases, number of new recoveries, and number of new deaths) are more readily available. In this article, a new methodology is proposed to estimate time-since-infection dependent recovery and death rates using easily available data sources, accommodating irregular data collection timings reflective of real-world reporting practices. The Nadaraya-Watson estimator is utilized to derive the number of new infections. This model improves the accuracy of epidemic progression descriptions and provides clear insights into recovery and death distributions. The proposed methodology is validated using COVID-19 data and its general applicability is demonstrated by applying it to some other diseases like measles and typhoid.


\vspace{0.5cm}


\medskip

\setcounter{equation}{1}
\setcounter{section}{0}
\setcounter{page}{1}


\section{Introduction}

In the literature of epidemiology, most of the existing compartmental epidemic models are based upon the assumptions of the classical susceptible-infected-removed (SIR) model (see, e.g., \cite{MR3197716} and a few relevant references therein). 
The classical SIR-type model (see \eqref{classical-SIR_model} and \eqref{r_n_d_n}) is a compartmental epidemic model based upon two main assumptions: (i) the disease incidence rate at time $t$ is proportional to the product of the number of susceptible $S(t)$ and the number of infected $I(t)$, (ii) the recovery and death rates at time $t$ are fixed for all infected individuals and assumed to be proportional to the number of infected individuals $I(t)$ at time $t$. The first assumption is justified for homogeneous population, but the second assumption has a limited applicability. 

In reality, for various infectious diseases, the time-to-recovery or time-to-death may significantly vary from one individual to another, and is determined by the individual's demographic variation, immune strength, and health condition due to the medical history \cite{bmb,clancy1999outcomes,lloyd2001destabilization}. Similarly, the infected individuals who died due to the disease, the time they spent under medical treatment is not fixed throughout the infected compartment. For instance, during the COVID-19 pandemic, several studies highlighted that the time from symptom onset to recovery or death varied significantly among patients, influenced by factors like age, pre-existing conditions, and severity of infection \cite{dessie2022assessment,Verity}. Similar variations in time-to-recovery or time-to-death have been documented for diseases such as measles, tuberculosis, typhoid, and malaria, where factors like HIV status, drug resistance, age, and presence of co-infections significantly affect recovery timelines \cite{terefe2018modeling,wesley1978immunological,typhoid_info}. Considering these variations, using a distribution function to represent the time-to-recovery or time-to-death in epidemic models can provide a more realistic and accurate representation of the disease dynamics within a population \cite{bmb}.

Further, in the context of vaccine preventable diseases (VPD), a critical question asked by the global public health and scientific research communities is: what is the minimum fraction of the unvaccinated susceptible population that must be immunized to halt an epidemic?  A well-established concept in the mathematical modeling of infectious diseases is the vaccine-derived herd immunity threshold (HIT) denoted by $p_c$, calculated using the formula $p_c=1 - \frac{1}{\mathcal{R}_0}$, where $\mathcal{R}_0$ represents the basic reproduction number, indicating the average number of secondary infections caused by a single infectious individual during their infectious period in a completely susceptible population. This vaccine-induced HIT is calculated for various diseases such as measles, smallpox, mumps, and rubella, as shown in Table-\ref{table_p_c}, using the formula $p_c=1 - \frac{1}{\mathcal{R}_0}$. Traditionally, $\mathcal{R}_0$ is determined using various SIR-type epidemic models, as exemplified in the classical SIR model (model-\eqref{classical-SIR_model}) formula given by $\mathcal{R}_0=\frac{\beta S_0}{N(r_0+d_0)}$. Here, the force of infection ($\beta$) is estimated by fitting the model with real incidence data, while $r_0$ and $d_0$ are chosen based on the average duration required for recovery and death, respectively. However, in reality, recovery and death rates may vary over time-since-infection. In the epidemiological literature, the selection of $r_0$ and $d_0$ typically relies on measures of central tendency such as mean, median, and mode, corresponding to the time-since-infection dependent recovery and death distributions $r(\eta)$ and $d(\eta)$ respectively. However, given the various factors or disease strains affecting $r_0$ and $d_0$, its time to reconsider the approach of determining these rates solely based on central tendency measures. This reevaluation is essential for advancing our understanding of disease progression and deciding various vaccination or intervention strategies for more precise and effective disease control.

\begin{table}[ht!]
    \centering
        \caption{Values of $\mathcal{R}_0$ and vaccine-driven HIT ($p_c$)}
        \begin{tabular}{llllll}
        \toprule
         & Measles& Rubella & Mumps & Smallpox\\
        \midrule
      $\mathcal{R}_0$& 18 & 6&8 &2.3\\
       $p_c$& 94\% & 83\% & 87\% & 57\%\\
        \bottomrule
    \end{tabular}
    \label{table_p_c}
\end{table}


Existing studies indicate that assuming constant recovery and death rates ($r_0$ and $d_0$) can lead to an overestimation of actual recovery and death outcomes \cite{bmb,jmb}. In contrast, incorporating time-since-infection dependent recovery and death rates, $r(\eta)$ and $d(\eta)$, offers a more accurate description of the epidemic. However, the survey data from which we can estimate these rates are often unavailable. The probability of recovery or the probability of death for an infected individual at a particular time-since-infection can be estimated by collecting individual level disease data with dates of disease onset and recovery (or death). To the best of our knowledge, availability of individual level disease data is scarce. On the other hand, some epidemiological data like daily number of new infection ($J(t)$), active number of infected individuals ($I(t)$), daily number of new recovery ($R_{new}(t)$) or daily number of new deaths ($D_{new}(t)$) etc. may be relatively easily available \cite{worldometer}.

  In this work, we develop a method that helps to estimate $r(\eta)$ and $d(\eta)$ using easily available epidemiological data. Our proposed method also provides added flexibility in the accumulation of epidemiological data, as it does not require that all the epidemiological data used in the estimation process are recorded at the same time intervals such as days, months etc. For instance, data related to one aspect of the disease incidence might be reported at daily interval, while information on recovery may be available at the weekly interval. This variability in timings of the collected data often occurs, especially when individuals are not fully aware or conscious of the importance of timely reporting. In our proposed estimation process, we do not impose rigid constraints on the temporal alignment of the epidemiological data used, making our method particularly valuable in application to real-world situations where data collection may be fragmented or delayed. Furthermore, we validate our proposed method using available data and employed it to obtain the explicit forms of $r(\eta)$ and $d(\eta)$. This, in turn, aids in determining basic reproduction number ($\mathcal{R}_0$) and the herd immunity threshold ($p_c$) more accurately for each of the examples considered.

It is clear from the expression in the SIR model (see \eqref{eqn_rn_1}, \eqref{eqn_rn_2}, \eqref{distributed_model}), the estimation of the function $J(.)$ is inevitable to carry out the study. In the context of estimating a function, there are a handful number of techniques available in the statistics literature (see, e.g., \cite{MR1383587}) for time series or ordinal data. Among them, one of the most well-known estimator is the Nadaraya-Watson type estimator (see, e.g., \cite{MR0172400} and \cite{MR0185765}), which is easy to compute for a given data, along with good practical and theoretical interpretations as well (see \cite{MR1383587}). Motivated by its simple and wide interpretability, in this work, we use the Nadaraya-Watson type estimator based on the data $(t_{i}, J_{i})_{i = 1}^{n}$, where $t_{i}$ is the $i$-th time point on $[0, T]$, $J_{i}$ is the value of $J(t)$ at $t = t_{i}$, and $n$ is the sample size.

 The Nadaraya-Watsdon type estimator \cite{MR1383587} is simply a weighted average of the responses (here $J_{i}$ for estimating $J(.)$), where the weights depend on the choice of a kernel function along with a tuning parameter named bandwidth. The main task of the kernel function is to control the smoothness/curvature of the estimated function, whereas the bandwidth controls the spread of the estimated function. For a given problem, as the choices of both the bandwidth and the kernel are decided by the user only, the use of Nadaraya-Watson type estimator is much flexible in practice. Besides, from the theoretical point of view, since the Nadaraya-Watson estimator is a certain weighted average of responses, one can express it as the unique minimizer of the corresponding weighted least squares problem (see, e.g., \cite{MR4511147} and a few references therein), which is also well-known as the local constant estimator in the literature of non-parametric regression (see, e.g., \cite{MR1383587}). Given the variety of existing methods for solving optimization problems related to weighted least squares (see, e.g., \cite{MR1920390}), the Nadaraya-Watson estimator offers a distinct theoretical advantage. Moreover, in view of the fact that the  Nadaraya-Watson estimator is essentially the local constant estimator of the regression function, one may be motivated to consider the techniques of local polynomial regression with a certain degree of polynomial, and in that case, it is possible to have the estimator of the $p (\geq 1)$-th order derivative of $J(.)$ when the degree of polynomial is $p$ is used to approximate the unknown curve $J(.)$. However, here it should be pointed out that the variability of the estimators will increase as the degree of the polynomial $p$ increases, and hence, there is a trade off between the choice of $p$ and the variability of the estimators of $l$-th order derivative of $J(.)$ for all $l\in\{1, \ldots, p\}$. 

The rest of the article is organized as follows. In Section \ref{section-2}, we discuss the methodology of the paper, including the materials, methods, and the estimation procedure. The results and main findings are discussed in Section \ref{section-3}.




\section{Methodology}\label{section-2}
\subsection{Materials and methods}

{\bf Classical SIR-type model and the model with time-since-infection dependent recovery and death rates:} 

\vspace{0.1in} 

\noindent Suppose that  $S(t)$, $I(t)$, $R(t)$ and $D(t)$ denote the number of susceptible, infected, recovered and dead individuals at time $t$, then the classical SIR-type model can be written as follows:
\begin{subequations}\label{classical-SIR_model}
    \begin{eqnarray}
    \frac{dS(t)}{dt} &=& -\frac{\beta}{N}S(t) I(t),\;\;
    \frac{dI(t)}{dt} = \frac{\beta}{N}S(t) I(t)-R_{new}(t)-D_{new}(t),\\
   && \frac{dR(t)}{dt} = R_{new}(t),\;\;\; \frac{dD(t)}{dt} = D_{new}(t),
    \end{eqnarray}
\end{subequations}
where $\beta$ is the transmission rate, $N$ is the total population size, $R_{new}(t)$ and $D_{new}(t)$ are the daily new recovery and deaths with constant rates $r_0$ and $d_0$ respectively, and 
\begin{equation}\label{r_n_d_n}
    R_{new}(t)=r_0 I(t),\;\; D_{new}(t)=d_0I(t).
\end{equation} 

Now, let $J(t)$ denote the number of new infections at time $t$. Hence, we have  
$$J(t)=\frac{\beta S(t) I(t)}{N},\;\; \text{and} \;\; \frac{dS(t)}{dt}=-J(t).$$
Suppose that $r(\eta)$ and $d(\eta)$ denote the probability distributions of recovery and death when the time-since-infection for an infected individual is $\eta$. Consequently,  the number of daily new recovery and new death at time $t$ are,  respectively, given by\\
\begin{subequations}
  \begin{tabularx}{\textwidth}{Xp{0.05cm}X}
\begin{equation}\label{eqn_rn_1}
R_{new}(t)=\int_0^t r(t-\eta) J(\eta) d\eta,
\end{equation}
& &
\begin{equation}\label{eqn_rn_2}
D_{new}(t)=\int_0^t d(t-\eta) J(\eta) d\eta.
\end{equation}
 \end{tabularx}
\end{subequations}
Hence, the modified model with time-distributed recovery and death rates will be : 
\begin{subequations}\label{distributed_model}
    \begin{eqnarray}
    \frac{dS(t)}{dt} &=& -J(t),\\
    \frac{dI(t)}{dt} &=& J(t)-R_{new}(t)-D_{new}(t),\\
    \frac{dR(t)}{dt} &=& R_{new}(t)=\int_0^t r(t-\eta) J(\eta) d\eta,\\
    \frac{dD(t)}{dt} &=& D_{new}(t)=\int_0^t d(t-\eta) J(\eta) d\eta.
    \end{eqnarray}
\end{subequations}
In \cite{bmb}, it was shown that the model \eqref{distributed_model} with \eqref{eqn_rn_1} and \eqref{eqn_rn_2} can capture the epidemic progression more accurately as compared to the model \eqref{classical-SIR_model} with \eqref{r_n_d_n}.\\

\noindent {\bf Basic reproduction number for the model \eqref{distributed_model}:}\; 

\vspace{0.1in}

\noindent We assume that in the beginning of an epidemic, $S(t) \approx S_0$ and $I(t) \approx 1$. For a single infected individual, if the individual has infectious period $\eta$, with the probability $r(\eta)+d(\eta)$, the total number of secondary infections then would be 
$$ \frac{\beta S_0}{N} \times \eta \times \left(r(\eta)+d(\eta)\right).$$

Now, suppose that the latent period of the infection is $\tau$ days, and afterwards,  integrating the above quantity over all possible infectious period $\eta > \tau$, we have the average number of secondary infections caused by a single infective as follows:
\begin{equation}\label{r0_distributed}
\mathcal{R}^1_0=\frac{\beta S_0}{N} \int_{\tau}^{\infty} \eta \;(r(\eta)+d(\eta)) d\eta,
\end{equation}
and hence, $p^1_c=1-\frac{1}{\mathcal{R}^1_0}$. Furthermore, for a more realistic scenario, if we assume that $\beta$ depends on the time-since-infection $\eta$, i.e., $\beta \equiv \beta(\eta)$, then the expression of basic reproduction number can be modified as:
\begin{equation}\label{r0_distributed_beta}
\mathcal{R}^1_0=\frac{ S_0}{N} \int_{\tau}^{\infty} \beta(\eta) \eta \;(r(\eta)+d(\eta)) d\eta.
\end{equation}
These expressions of $\mathcal{R}^1_0$ give more accurate estimation of the basic reproduction number, and as a result, we have a more accurate estimate of $p^1_c$. Moreover, observe that calculating $\mathcal{R}^1_0$ as in both the formula \eqref{r0_distributed} and \eqref{r0_distributed_beta} need the explicit forms of $r(\eta)$ and $d(\eta)$. This fact motivated us to set a goal in this work to estimate these $r(\eta)$ and $d(\eta)$ with the help of the available epidemiological data.\\



\noindent {\bf Nadaraya-Watson estimator:}\; 

\vspace{0.1in}
In the estimation methodology we use the Nadaraya-Watson estimator which is defined as follows: Suppose that we have a data set consisting of $n$ pairs of observations $(x_i, y_i)$, $i=1,2, \cdots, n$ on (X, Y), where $x_i$ is the input or predictor variable, and $y_i$ is the output or response variable. Now, let $E(Y| X = x) := m(x)$, where $m$ is an unknown function.
Then, in order to estimate $m(.)$, the estimator proposed by Nadaraya and Watson (see \cite{MR0172400} and \cite{MR0185765}) at a specific point $\xi$ is defined as follows:
\begin{equation}\label{nadaraya_defn}
\hat{m}_{n}(\xi)=\frac{ \sum\limits_{i=1}^{n} l\left(\frac{\xi-x_i}{h_n}\right) y_i}{ \sum\limits_{i=1}^{n} l\left(\frac{\xi-x_i}{h_n}\right)},
\end{equation}
where $l(.)\geq 0$ is a kernel function satisfying $\int l(x)dx = 1$, with is a positive sequence of bandwidth $\{h_n\}_{n\geq 1}$. In order to have some technical and practical advantage, in this study, we consider  $h_n=n^{-1/5}$, and $l(x)=\frac{1}{\sqrt{2 \pi}}e^{-\frac{x^2}{2}},$ for $x \in \mathbb{R}$ throughout the article, unless mentioned earlier. Regarding the choices of the bandwidth and the kernel function, the readers are referred to \cite{MR0848134}.\\

\noindent {\bf Distributional assumptions based on epidemiology:}\; 

\vspace{0.1 in}

\noindent In this work, we consider the following epidemiologically justified assumptions:
\begin{itemize}

\item We assume that the recovery (or death) distributions for individuals who will recover (or die) follow the gamma probability density function (pdf) as given below
\begin{equation}\label{gamma_pdf_defn}
    f(t; a, b) = \frac{1}{\Gamma(a)b^a} x^{a-1} e^{-\frac{t}{b}},
\end{equation}
where, $a>0$ is the shape parameter and $b>0$ is the scale parameter.
This choice of the gamma distribution agrees with the observed pattern where the probability of recovery  for an infected individual is initially low just after infection, gradually increases as the immune response strengthens, peaks at the height of immunity, and then decreases. The use of a gamma distribution to describe recovery and death distributions is a well known fact in the field of epidemiology \cite{bailey1954statistical,chowell_book}.

\item We assume that the gamma distributions for time-distributed recovery and death rates have shape parameters $a>1$. This ensures the probability density function rises to a peak and then falls, reflecting the observed patterns of recovery and death.

\item The mode of a gamma-distributed recovery (or death) rate has epidemiological significance because it represents the most likely duration for individuals to recover (or die) from an infectious disease. We assume that the mode of the gamma distributions for the recovery and death rates vary between a feasible range of $T^l$ and $T^u$, i.e., $T^l \leq mode= (a-1)b \leq T^u$. 

\end{itemize}

\subsection{ Estimation procedure for $r(\eta)$}

\vspace{0.1in}

\noindent Let us first recall $r(\eta)$ from \eqref{eqn_rn_1}, and the estimation of $r(.)$ is as follows. Throughout this study, we assume that $t\in[0, T]$, where $T > 0$. Suppose that the real data for $J(t)$ (see \eqref{distributed_model}) is available at time points $t=t_1, t_2, \cdots, t_{n}$ and denote $J_{i} = J(t_{i})$ for $i = 1, \ldots, n$. In view of Nadaraya-Watson estimator (see \eqref{nadaraya_defn}), for any $\xi \in [0,\; T]$, we define,
\begin{equation}\label{jhat_1}
\widehat{J}(\xi)=\frac{ \sum_{i=1}^{n} l(\frac{\xi-t_i}{h_n}) J_i}{ \sum_{i=1}^{n} l(\frac{\xi-t_i}{h_n})},
\end{equation}
where, $h_n=n^{-1/5}$, and $l(x)=\frac{1}{\sqrt{2 \pi}}e^{-\frac{x^2}{2}},$ for $x \in \mathbb{R}$.
Now, assume that $p_0$ is the survival probability of infected individuals corresponding to a specific infection,  and $f_r(t)$ is the gamma pdf describing the probability of recovery as a function of the time-since-infection for an infected individual who will recover in future. Hence, one can write $$r(t) = p_0 f_r(t),\;\;
\text{with} \;\;f_r(t)=\frac{1}{b_r^{a_r} \Gamma (a_r)} t^{a_r-1} e^{-\frac{t}{b_r}}, $$
where, the shape parameter $a_r$ is assumed to be greater than 1,  and $b_r>0$ is the scale parameter.

Next, define
\begin{equation}\label{Rncap_1}
\widehat{R}_{new}(t;a_r,b_r):=\int_0^t p_0 f_r(t-\eta) \widehat{J}(\eta) d \eta,\;\;\;\;\text{for} \;\;t \in [0, T].
\end{equation}
Suppose that we have real data of daily number of new recovery $R_{new}$, at the time points $\widetilde{t}_1, \widetilde{t}_2, \cdots, \widetilde{t}_{m},$ and the data are represented by $\widetilde{R}_{{new}_1}, \widetilde{R}_{{new}_2}, \cdots, \widetilde{R}_{{new}_m}.$ Note that, in principle, the two sets 
$$\big\{ \widetilde{t}_1, \widetilde{t}_2, \cdots, \widetilde{t}_{m} \big\} \;\; \text{and} \;\; \big\{ t_1, t_2, \cdots, t_n \big\}$$
may not be equal; however, the domains of both sets are the same, i.e., $[0, T]$. Afterwards, suppose that the shape parameter $a_r$ and scale parameters $b_r$ varies in the feasible region $[0, A]$ and $[0, B]$ respectively, for some positive constants $A$ and $B$. Define the function
$$E_r:[0, A] \times [0, B] \to \mathbb{R}_{\geq0}$$
by
\begin{equation}\label{Er_1}
E_r(a_r,b_r)=\sum_{j=1}^{m} \Big( \widehat{R}_{new}(\widetilde{t}_j;a_r,b_r)-\widetilde{R}_{{new}_j} \Big)^2,\;\; \forall (a_r,b_r) \in [0, A] \times [0, B], 
\end{equation} where $\widehat{R}(.; ., )$ is the same as defined in \eqref{Rncap_1}. 
The objective is now to minimize the function $E_r(a_r,b_r)$ in the region $ [0, A] \times [0, B]$. Observe that $E_r$ is a continuous function on a compact set $[0, A] \times [0, B]$, and hence, it attains its global minimum in the domain $[0, A] \times [0, B]$.
 Let $E_r$ attain its minima at $(a^*_r, b^*_r)$ and denote 
 \begin{equation}\label{eqb_optimal_r}
  r^*(t) = p_0 f^*_r(t),
  \end{equation}
with
\begin{equation}\label{eqb_optimal_r1}
f^*_r(t)=\frac{1}{{(b^*_r)}^{a^{*}_r} \Gamma (a^{*}_r)} t^{a^{*}_r-1} e^{-\frac{t}{b^{*}_r}}.
\end{equation} 
Hence, finally, the estimator for $r(t)$ is $r^*(t)$ as in formula \eqref{eqb_optimal_r}.


\vspace{0.1in}

Algorithm \ref{alg:rn} shows how to estimate $r(\eta)$. Similar method for estimation of $d(\eta)$ is discussed in the supplementary materials.
\begin{algorithm}
    \SetAlgoLined
    \SetKwInOut{Input}{Input}\SetKwInOut{Output}{Output}
    \Input{Data $\{t_i\}$, $\{J_i\}$, $\{\widetilde{t}_j\}$, $\{\widetilde{R}_{{new}_j}\}$, where $i=0, 1, \cdots, n$, $j=0, 1, \cdots, m$ }
    \Output{$(a^*_r, b^*_r)$ and $r^*(t)$}

    \nl Define time interval $[0, T]$.

    \nl Calculate the Nadaraya-Watson Estimator $\widehat{J}(\xi)$ using the formula \eqref{jhat_1}.

    \nl Define the survival probability $p_0$.

    \nl Define the feasible range of the shape ($a_r$) and scale ($b_r$) parameters, i.e., $(a_r,b_r) \in [0,A]\times [0,B]$.

    \nl Define $T_r^l$ and $T_r^u$.

    \nl Create a mesh grid for the shape ($a_r$) and scale ($b_r$) parameters in the domain $[0, A]\times [0,B]$.

    \nl \For {each combination of $(a_r,b_r)$ with $T_r^l \leq (a_r-1)b_r \leq T_r^u$}{

        \nl Calculate $\widehat{R}_{new}(t;a_r,b_r)$ as defined in \eqref{Rncap_1}.

        \nl Calculate the error term $E_r(a_r,b_r)$ as defined in \eqref{Er_1}.
        
    }

    \nl Calculate the minimum of $E_r$ in the whole mesh and denote the corresponding mesh grid by $(a_r^*,b_r^*)$.

    \nl Compute $f_r^*(t)$ and $r^*(t)$ using the formula \eqref{eqb_optimal_r1} and \eqref{eqb_optimal_r}.
\caption{Estimation of $r(\eta)$}
\label{alg:rn}
\end{algorithm}

\section{Results and Findings}\label{section-3}

We validate our proposed methodology and explore its applicability using epidemiological data, as detailed in the supplementary materials.\\

\noindent {\bf Model validation:}\; 

\vspace{0.1in}

\noindent In this section, we validate the proposed methods through real data examples. In Example 1, the publicly available survey data from \cite{survey_data,Verity} is used. In this survey individual-cases with date of disease onset and the date of recovery (or death) for patients who recovered (or died) were collected in the context of COVID-19 from Hubei, mainland China in the period 23 January, 2020 to 15 April, 2020. The data from $120$ recoveries and $31$ deaths from \cite{survey_data,Verity} were used to estimate the probabilities of recovery and death, respectively (see Fig.~\ref{bar_1}). The survival probability was estimated to be  $p_0=0.97$ from the Covid-19 data available in \cite{worldometer}. 
\begin{figure}[ht!]
\begin{center}
\mbox{
\subfigure[]{\includegraphics[scale=0.4]{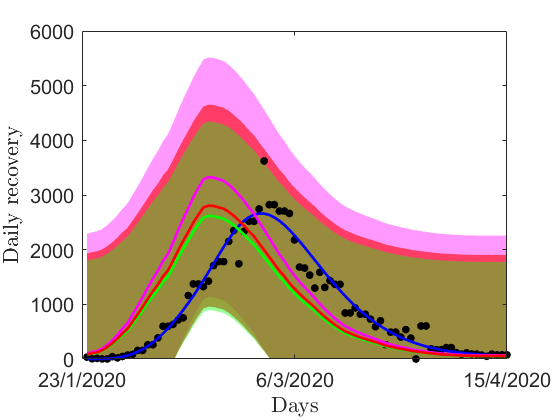}}
\subfigure[]{\includegraphics[scale=0.4]{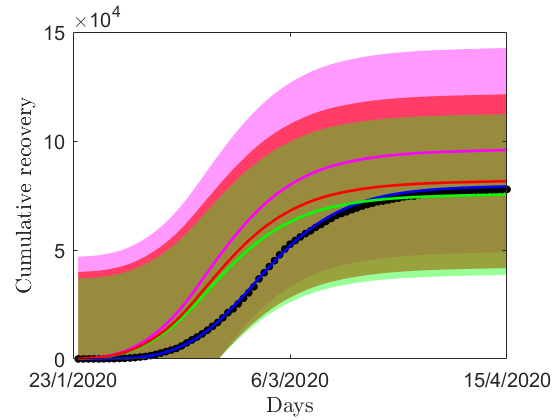}}}
\mbox{
\subfigure[]{\includegraphics[scale=0.4]{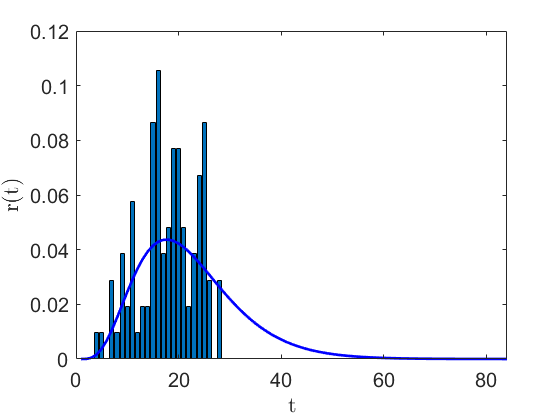}}
}
\caption{The black dots correspond to real data for China and the blue bars are the survey data of onset-to-recovery probability. The blue curves in panel (a), (b) and (c) correspond to $\widehat{R}_{new}(t;a^*_r,b^*_r)$ in formula \eqref{Rncap_1}, the cumulative $\widehat{R}_{new}(t;a^*_r,b^*_r)$, and the $r^*(t)$ respectively, with the optimal situation $(a^*_r,b^*_r)=(4.7,4.5)$. The daily recovery from the classical SIR-type model is given by $r_0I(t)$, where $I(t)$ is the active cases at time $t$, and $r_0=p_0/ mean$ (green curve); $r_0=p_0/ median$ (red curve); $r_0=p_0/ mode$ (magenta curve). The 3-sigma range for the green, red and magenta curves are shown by the shaded region with corresponding transparent colors.}
\label{validate_r}
\end{center}
\end{figure}

\begin{figure}[ht!]
\begin{center}
\mbox{
\subfigure[]{\includegraphics[scale=0.4]{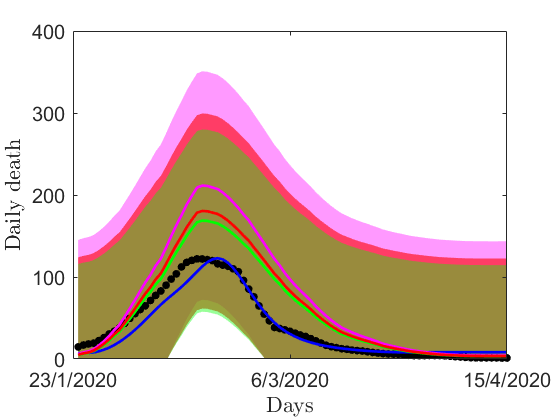}}
\subfigure[]{\includegraphics[scale=0.4]{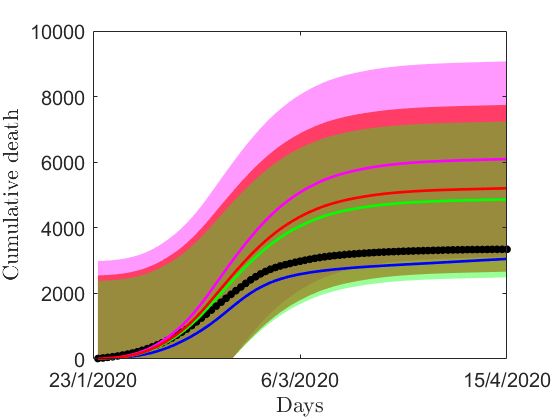}}}
\mbox{
\subfigure[]{\includegraphics[scale=0.4]{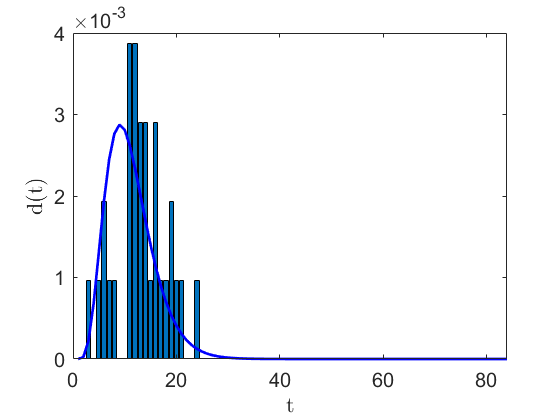}}}
\caption{The black dots correspond to real data for China and the blue bars are the survey data of onset-to-death probability. The blue curves in panel (a), (b) and (c) correspond to $\widehat{D}_{new}(t;a^*_d,b^*_d)$ in formula \eqref{Dncap_1}, the cumulative $\widehat{D}_{new}(t;a^*_d,b^*_d)$, and the $d^*(t)$ respectively, with the optimal situation $(a^*_d,b^*_d)=(4.95,2.05)$. The daily death from the classical SIR-type model is given by $d_0I(t)$, where $I(t)$ is the active cases at time $t$, and $d_0=(1-p_0)/ mean$ (green curve); $d_0=(1-p_0)/ median$ (red curve); $d_0=(1-p_0)/ mode$ (magenta curve). The 3-sigma range for the green, red and magenta curves are shown by the shaded region with corresponding transparent colors.}
\label{validate_d}
\end{center}
\end{figure}
The number of daily and cumulative recoveries (deaths) were calculated using the time varying estimator $$R_{new}(t)=\int_0^t r(t-\eta) J(\eta)d \eta.$$ and the estimator from the classical SIR model $$R_{new}(t)=r_0 I(t).$$ In the classical estimator $r_0$ was estimated by the three main measures of central tendency, i.e.,  $r_0=p_0/mean$, or $r_0=p_0/median$ or $r_0=p_0/mode$. The estimation results from the time varying and classical models were then compared to show the superiority of the proposed approach.

The blue curve in Fig.~\ref{validate_r}c represent the estimated recovery distribution $r^*(t)$ given in \eqref{eqb_optimal_r} with $(a^*_r,b^*_r)=(4.7,4.5)$. The bars show the probability of recoveries obtained from the survey data as discussed above. The corresponding gamma distribution $f_r^*$ in \eqref{eqb_optimal_r1} has mean $=21.15$, median $=19.64$ and mode $=16.65$. Using  $r^*(t)$ from Fig.~\ref{validate_r}c, the number of daily and cumulative recoveries are computed as shown in Fig.~\ref{validate_r}. The black dots in figures.~\ref{validate_r}a and ~\ref{validate_r}b correspond to the actual daily recoveries and cumulative recoveries, respectively, while the corresponding estimated values are shown using the blue lines.  The green, red and magenta curves in Fig.~\ref{validate_r}a correspond to $R_{new}(t)=r_0 I(t)$, with $r_0=p_0/mean\approx 0.0459$, $r_0=p_0/median \approx0.0494 $ and $r_0=p_0/mode \approx 0.0583$, respectively. Similar  symbols and colors in Fig.~\ref{validate_r}b represent the estimated cumulative recoveries. Also, 3-sigma intervals for the classical estimator (mean, median and mode) are shown by the shaded region with corresponding transparent colors.
\begin{figure}[ht!]
\begin{center}
\mbox{
\subfigure[]{\includegraphics[scale=0.35]{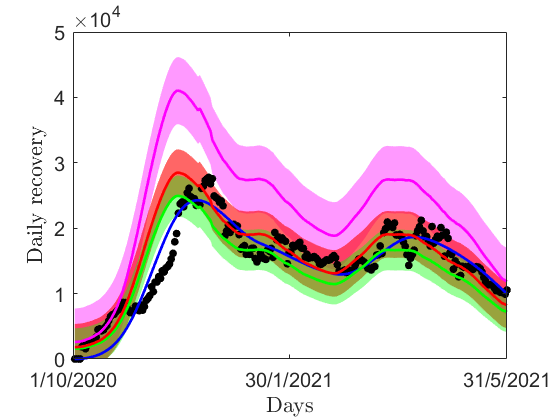}}
\subfigure[]{\includegraphics[scale=0.35]{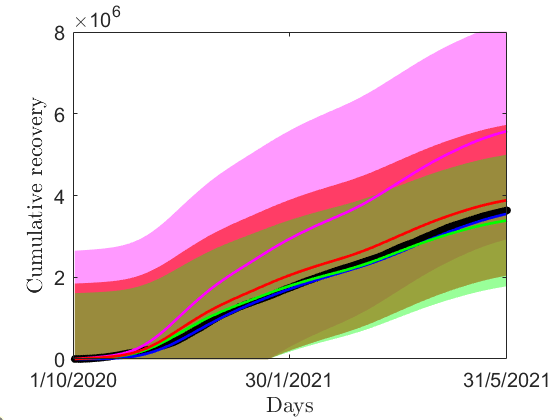}}}
\mbox{
\subfigure[]{\includegraphics[scale=0.35]{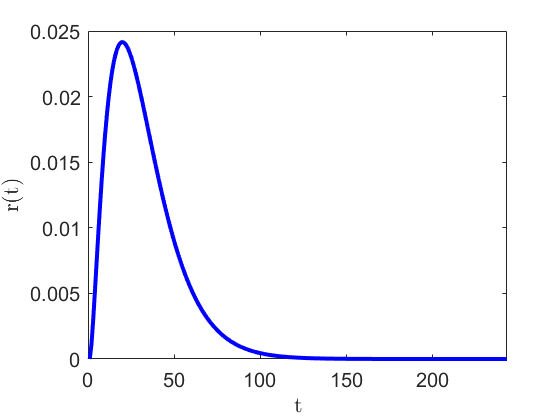}}
}
\caption{The black dots correspond to real data for COVID-19 in Italy during 1/10/2020 to 31/5/2021, when the Alpha strain was dominant. The blue curves in panel (a), (b) and (c) correspond to $\widehat{R}_{new}(t;a^*_r,b^*_r)$ in formula \eqref{Rncap_1}, the cumulative $\widehat{R}_{new}(t;a^*_r,b^*_r)$, and the $r^*(t)$ respectively, with the optimal situation $(a^*_r,b^*_r)=(2.55,12.2)$. The daily recovery from the classical SIR-type model is given by $r_0I(t)$, where $I(t)$ is the active cases at time $t$, and $r_0=p_0/ mean$ (green curve); $r_0=p_0/ median$ (red curve); $r_0=p_0/ mode$ (magenta curve). The 3-sigma range for the green, red and magenta curves are shown by the shaded region with corresponding transparent colors.
}
\label{italy_estimate_1_r}
\end{center}
\end{figure}

\begin{figure}[ht!]
\begin{center}
\mbox{
\subfigure[]{\includegraphics[scale=0.35]{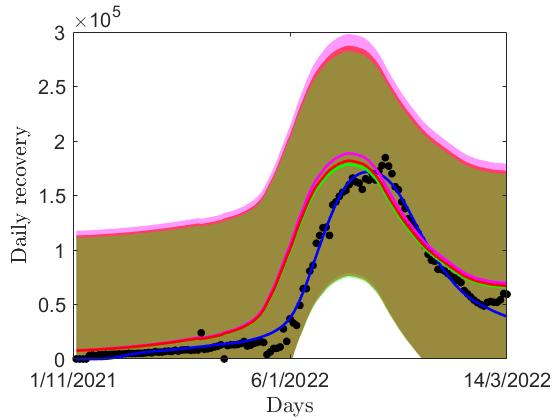}}
\subfigure[]{\includegraphics[scale=0.35]{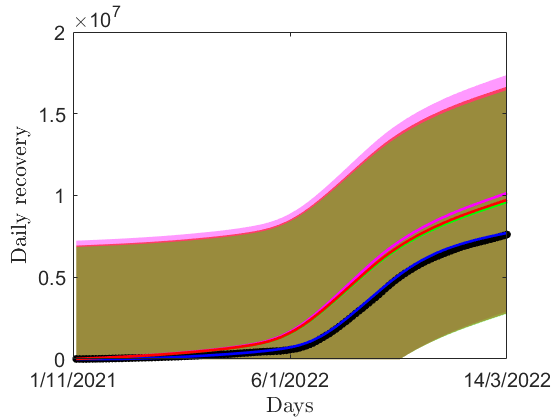}}}
\mbox{
\subfigure[]{\includegraphics[scale=0.35]{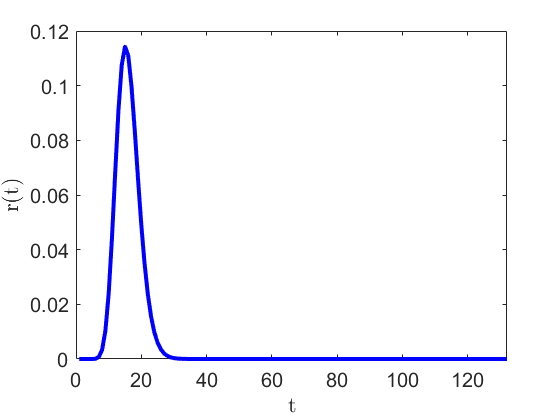}}
}
\caption{The black dots correspond to real data for COVID-19 in Italy during 1/11/2021 to 14/3/2022, when the Delta strain was dominant. The blue curves in panel (a), (b) and (c) correspond to $\widehat{R}_{new}(t;a^*_r,b^*_r)$ in formula \eqref{Rncap_1}, the cumulative $\widehat{R}_{new}(t;a^*_r,b^*_r)$, and the $r^*(t)$ respectively, with the optimal situation $(a^*_r,b^*_r)=(18.7,0.8)$. The daily recovery from the classical SIR-type model is given by $r_0I(t)$, where $I(t)$ is the active cases at time $t$, and $r_0=p_0/ mean$ (green curve); $r_0=p_0/ median$ (red curve); $r_0=p_0/ mode$ (magenta curve). The 3-sigma range for the green, red and magenta curves are shown by the shaded region with corresponding transparent colors.}
\label{italy_estimate_2_r}
\end{center}
\end{figure}

\begin{figure}[ht!]
\begin{center}
\mbox{
\subfigure[]{\includegraphics[scale=0.35]{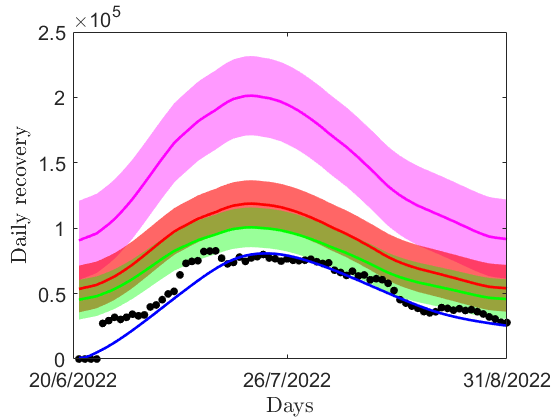}}
\subfigure[]{\includegraphics[scale=0.35]{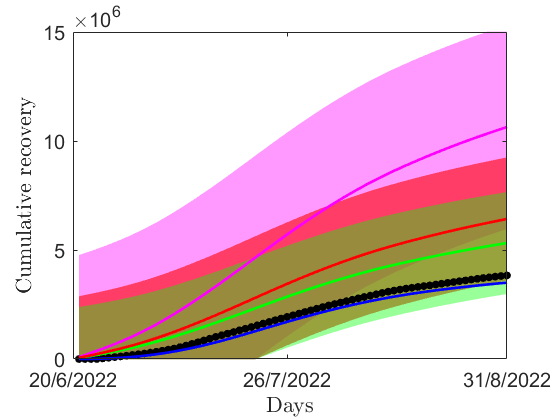}}}
\mbox{
\subfigure[]{\includegraphics[scale=0.35]{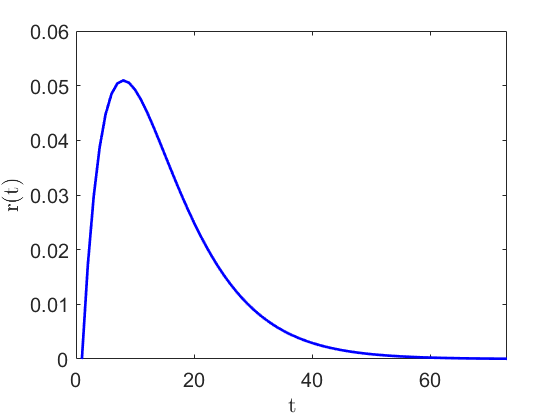}}
}
\caption{The black dots correspond to real data for COVID-19 in Italy during 20/6/2022 to 31/8/2022, when the Omicron strain was dominant. The blue curves in panel (a), (b) and (c) correspond to $\widehat{R}_{new}(t;a^*_r,b^*_r)$ in formula \eqref{Rncap_1}, the cumulative $\widehat{R}_{new}(t;a^*_r,b^*_r)$, and the $r^*(t)$ respectively, with the optimal situation $(a^*_r,b^*_r)=(2.01,6.98)$. The daily recovery from the classical SIR-type model is given by $r_0I(t)$, where $I(t)$ is the active cases at time $t$, and $r_0=p_0/ mean$ (green curve); $r_0=p_0/ median$ (red curve); $r_0=p_0/ mode$ (magenta curve). The 3-sigma range for the green, red and magenta curves are shown by the shaded region with corresponding transparent colors.}
\label{italy_estimate_3_r}
\end{center}
\end{figure}

From Fig.~\ref{validate_r}, we note that the estimated $r(t)$ using our time varying method agrees quite well with the survey data. The corresponding SSE value for daily recoveries obtained using our estimated recovery distribution $r(t)$ is $1.746 \times 10^3$. 
In contrast, the SSE corresponding to the classical estimator using the central tendency measures (mean, median, and mode) are $6.0967 \times 10^3$, $6.3177 \times 10^3$, and $7.3349 \times 10^3$, respectively.  This shows the drawback in using the classical SIR models to estimate the number of recoveries when the time to recovery varies for a disease. 

In Fig.~\ref{validate_d}, we note similar outcome for the death rate distribution $d(t)$. The estimated values of the shape and scale parameters for the death rate distribution is $(a^*_d,b^*_d)=(4.95,2.05)$. The estimated $d^*(t)$ with $(a^*_d,b^*_d)=(4.95,2.05)$ gives reasonably good estimate to the survey data (see Fig.~\ref{validate_d}c). We observe that the blue curves fits the real data of daily death and cumulative death. The corresponding SSE value for daily recoveries obtained using our estimated recovery distribution $r(t)$ is $117.2664$, whereas, none of the green, red and magenta curves fits the real data (SSE values for daily recoveries: $368.7897$ (green); $457.5948$ (red); $665.3864$ (magenta)).

Moreover, we observe that the green, red and magenta curves have a tendency to overestimate the number of deaths and also recoveries, which in turn may lead to an underestimation of the number of daily infections. Thus, the choice of $r_0$ and $d_0$ as used in the classical literature based upon the measures of central tendencies like, mean, median, mode, may not be effectively describing the epidemic progression.\\

\noindent {\bf Strain-specific recovery and death distributions:}\; 

\vspace{0.1in}

\noindent The impact of different viral strains on the immune system may differ leading to a variation in the disease progression. For instance, in the case of COVID-19, it was observed that the Delta variant exhibited more robust growth within lung tissues compared to the upper respiratory tract. Conversely, the Omicron variant showed diminished growth in lung tissues but thrived in the upper respiratory tract \cite{chan2021sars}. The Delta variant, with its heightened presence in lung tissues, has been associated with more extensive damage, leading to increased disease severity, whereas, the Omicron variant has milder disease severity. Thus, the time-since-infection dependent recovery and death rates are different for different strains. 

To see the effect of different Covid-19 strains, mainly Alpha, Delta and Omicron, on the disease progression, we use COVID-19 data from Italy in three different time periods when these strains were dominant. The proportion of different strains during the COVID-19 epidemic can be viewed in \cite{strain_specific_data}. For Italy, we consider three time periods: 1/10/2020 to 31/5/2021 (dominant strain is Alpha); 1/11/2021 to 14/3/2122 (dominant strain is Delta); 20/6/2022 to 31/8/2022 (dominant strain is Omicron). The corresponding estimation results for the recovery distribution $r(t)$ are shown in the Fig.~\ref{italy_estimate_1_r}, Fig.~\ref{italy_estimate_2_r} and Fig.~\ref{italy_estimate_3_r} for the three stains, respectively. Also, the estimated values of $(a^*_r, b^*_r)$ along with the mean, median and mode of the corresponding estimated gamma distributions are reported in Table.~\ref{table:strain_specific_r}.

\begin{table}[ht!]
    \centering
        \caption{Estimated recovery distribution during different dominant strains in Italy}
        \begin{tabular}{lrrrrr}
        \toprule
         Dominant strain & Estimated value  & Mean of $f^*_r$ & Median of $f^*_r$&Mode of $f^*_r$\\
           & of $(r^*_r,b^*_r)$ &  & &\\
        \midrule
       Alpha (Fig.~\ref{italy_estimate_1_r})& (2.55,12.2) & 31.11&27.23& 18.91\\
        Delta (Fig.~\ref{italy_estimate_2_r})& (18.7,0.8) & 14.96&14.76&14.16 \\
        Omicron (Fig.~\ref{italy_estimate_3_r})& (2.01,6.98) & 14.03 &11.71&7.05\\
        \bottomrule
    \end{tabular}
    \label{table:strain_specific_r}
\end{table}

\begin{figure}[ht!]
\begin{center}
\mbox{
\subfigure[]{\includegraphics[scale=0.28]{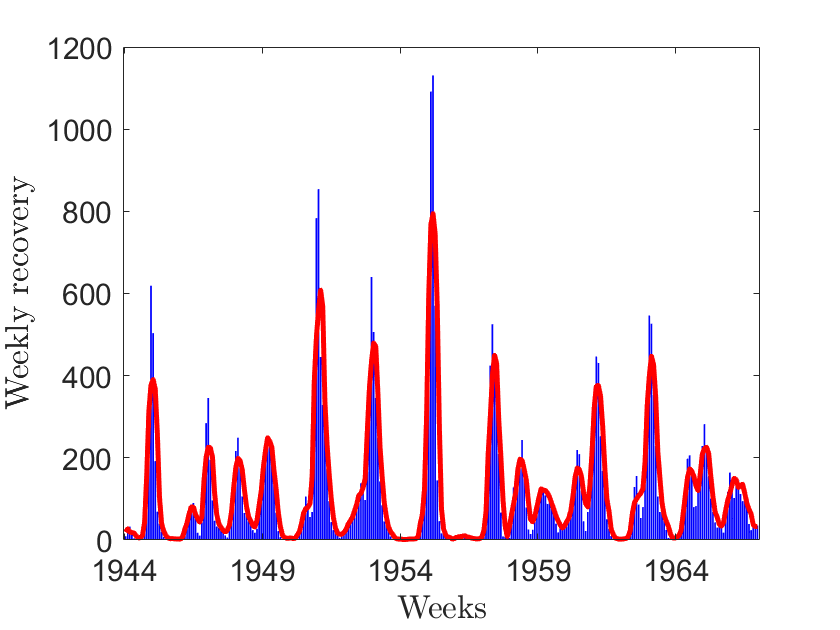}}
\subfigure[]{\includegraphics[scale=0.28]{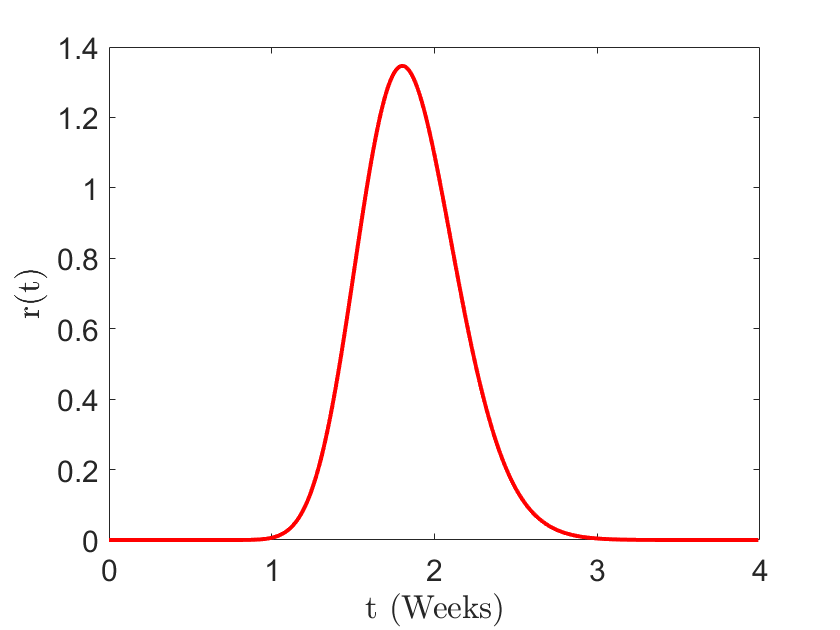}}}
\caption{The blue bars correspond to weekly real data for measles in Nottingham, UK from January, 1944 to December, 1966. The red curves in panel (a) and (b) correspond to $\widehat{R}_{new}(t;a^*_r,b^*_r)$ in formula \eqref{Rncap_1} and the $r^*(t)$ respectively, with $(a^*_r,b^*_r)=(38.16,0.0485)$.}
\label{measles_mumbai_1}
\end{center}
\end{figure}

\begin{figure}[ht!]
\begin{center}
\mbox{
\subfigure[]{\includegraphics[scale=0.28]{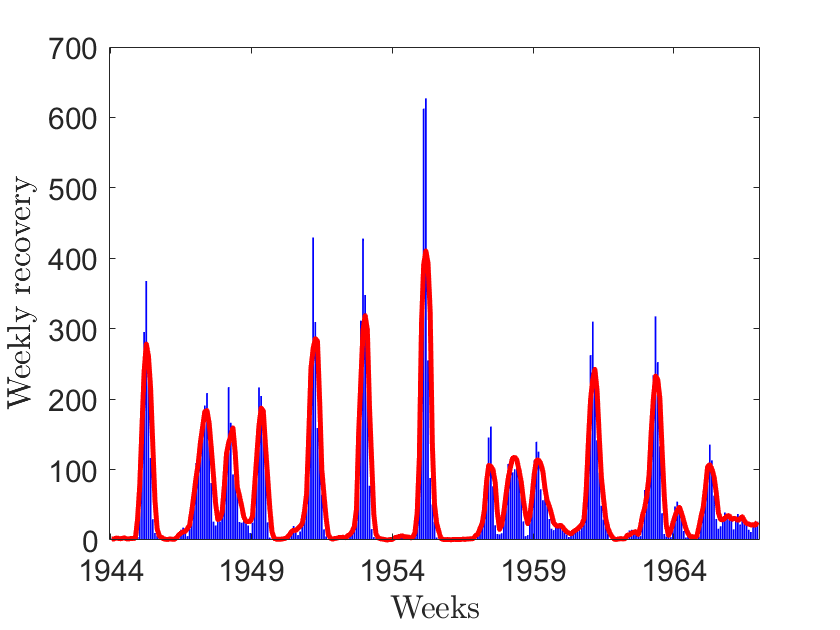}}
\subfigure[]{\includegraphics[scale=0.28]{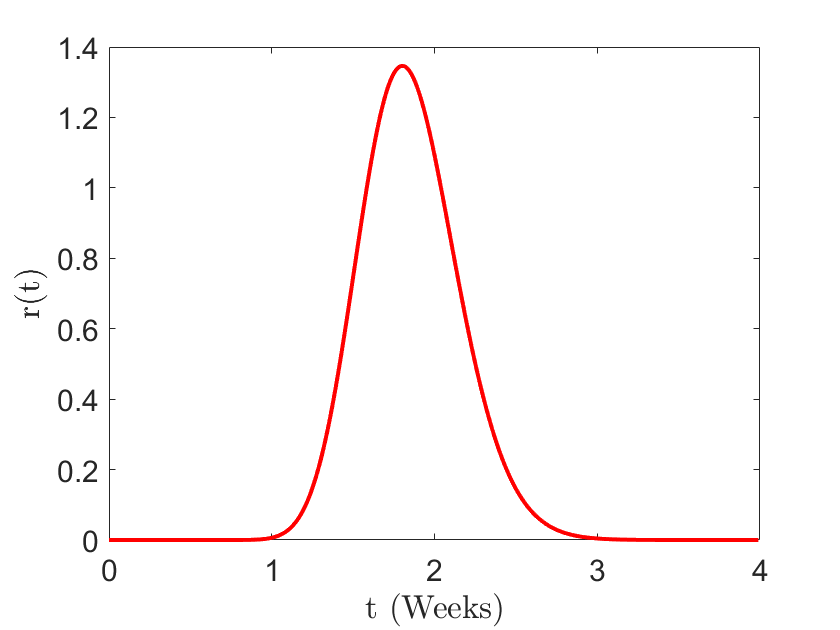}}}
\caption{The blue bars correspond to weekly real data for measles in Derby, UK from January, 1944 to December, 1966. The red curves in panel (a) and (b) correspond to $\widehat{R}_{new}(t;a^*_r,b^*_r)$ in formula \eqref{Rncap_1} and the $r^*(t)$ respectively, with $(a^*_r,b^*_r)=(38.16,0.0485)$.}
\label{measles_thane_1}
\end{center}
\end{figure}

\begin{figure}[ht!]
\begin{center}
\mbox{
\subfigure[]{\includegraphics[scale=0.4]{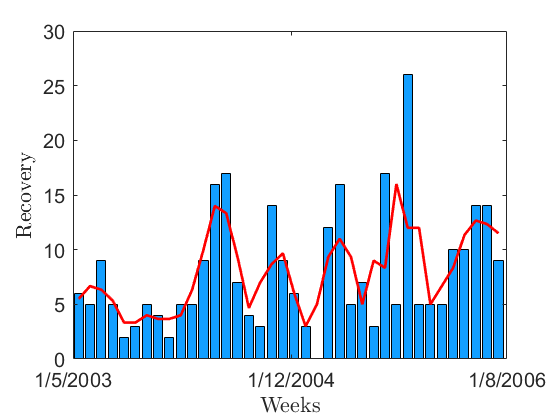}}
\subfigure[]{\includegraphics[scale=0.4]{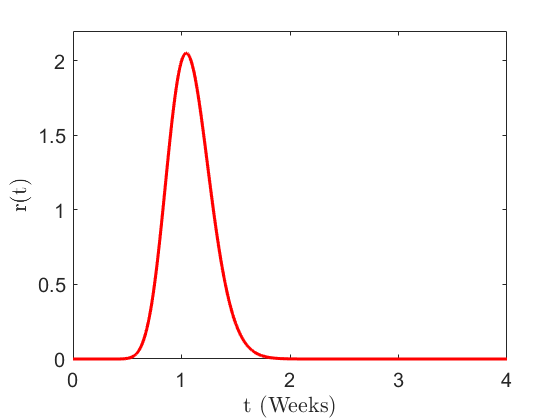}}}
\caption{The blue bars correspond to weekly real data for typhoid in Kolkata, India from May, 2003 to August, 2006. The red curves in panel (a) and (b) correspond to $\widehat{R}_{new}(t;a^*_r,b^*_r)$ in formula \eqref{Rncap_1} and the $r^*(t)$ respectively, with $(a^*_r,b^*_r)=(30,0.036)$.}
\label{typhoid_kolkata_1}
\end{center}
\end{figure}

From Fig.~\ref{italy_estimate_1_r}, Fig.~\ref{italy_estimate_2_r} and  Fig.~\ref{italy_estimate_3_r} for the three different strains, we observe that the blue curves fits the real data of daily and cumulative recovery much better than the green, red and magenta curves. For the Alpha strain, the green and red curves (which are corresponding to $r_0=p_0/mean$ and $r_0=p_0/median$, respectively) are close to the blue curve and the magenta curve (which is corresponding to $r_0=p_0/mode$) is far away from the real data. However, this is not the case for Delta and Omicron strain. The SSE values for each strain using the classical and time varying estimators are reported in Table~\ref{table:sse}. The overall point to note here is that when there is variation in the time to recovery and time to death for various strains of the disease, the time varying estimator performs better than the classical estimator. 
Similarly, the death distributions $d(t)$ during different dominant strains in Italy have been computed,  and the estimation results are given in the supplementary materials.\\

\begin{table}[ht!]
    \centering
    \caption{SSE values}
    \begin{tabular}{lcccc}
        \toprule
        Dominant strain & \multicolumn{4}{c}{SSE value for daily recoveries} \\
        \cmidrule(lr){2-5}
        & Blue  & Green  & Red  & Magenta  \\
        \midrule
        Alpha (Fig.~\ref{italy_estimate_1_r}) & $3.4401 \times 10^4$ & $5.7849 \times 10^4$ & $5.5168 \times 10^4$ & $1.3497 \times 10^5$ \\
        Delta (Fig.~\ref{italy_estimate_2_r}) & $7.8037 \times 10^4$ & $3.0543 \times 10^5$ & $3.1209 \times 10^5$ & $3.2776 \times 10^5$ \\
        Omicron (Fig.~\ref{italy_estimate_3_r}) & $6.5098 \times 10^4$ & $2.5436 \times 10^5$ & $3.6627 \times 10^5$ & $8.1476 \times 10^5$ \\
        \bottomrule
    \end{tabular}
    \label{table:sse}
\end{table}


{\bf Measles and typhoid data for some regions in UK and India.} We employed the same method of estimating $r(\eta)$ for some other diseases like measles and typhoid. We used the data of measles for Nottingham and Derby regions in UK from January, 1944 to December, 1966 (see  \cite{measles_data_UK}), and the data of typhoid is collected from \cite{kanungo2008epidemiology} for the region Kolkata in India from May, 2003 to August, 2006. The estimated time-since-infection dependent recovery distribution is shown in Fig.~\ref{measles_mumbai_1}, and Fig.~\ref{measles_thane_1} corresponding to the measles data for Nottingham and Derby regions, UK. The same for typhoid data obtained from Kolkata, India is shown in Fig.~\ref{typhoid_kolkata_1}. It is noteworthy that for measles, the central tendencies of the recovery distribution lie around 2 weeks (for Fig.~\ref{measles_mumbai_1}:  mean $=1.851$, median $=1.8389$, mode $=1.8023$, variance $=0.0898$; for Fig.~\ref{measles_thane_1}:  mean $=1.8508$, median $=1.8317$, mode $=1.8023$, variance $=0.0898$), whereas for typhoid, they lie around 1 week (for Fig.~\ref{typhoid_kolkata_1}:  mean $=1.08$, median $=1.0702$, mode $=1.0440$, variance $=0.0389$), that aligns with the realistic scenarios.

\section{Discussion}

In this work, we have developed a novel method to estimate the time-since-infection dependent recovery and death rates, $r(\eta)$ and $d(\eta)$, using readily available epidemiological data such as the daily number of new infections ($J(t)$), the active number of infected individuals ($I(t)$), the daily number of new recoveries ($R_{new}(t)$), and the daily number of new deaths ($D_{new}(t)$). Unlike traditional approaches, our method does not require that all epidemiological data be collected simultaneously. This flexibility is crucial in real-world scenarios where data collection can be fragmented or delayed, and it acknowledges the variability and often asynchronous nature of data reporting. Our method allows for the integration of data reported at different times, making it highly practical for situations where individuals may not report information promptly. By applying this method to available data, we successfully derived explicit forms of $r(\eta)$ and $d(\eta)$, which in turn facilitate a more accurate determination of important epidemiological metrics such as the basic reproduction number ($\mathcal{R}_0$) and the herd immunity threshold ($p_c$). This approach enhances the precision of epidemic modeling and supports more informed public health decision-making.

\begin{figure}[ht!]
\begin{center}
\mbox{
\subfigure[]{\includegraphics[scale=0.45]{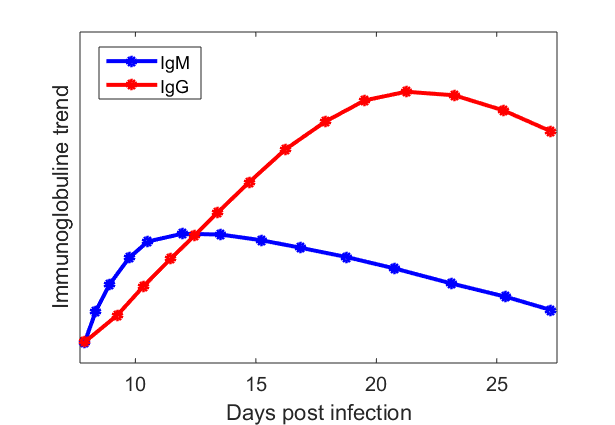}}
\subfigure[]{\includegraphics[scale=0.45]{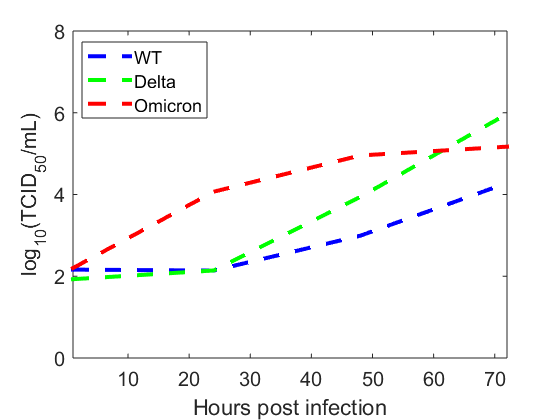}}}
\caption{(a) Trends in IgG and IgM profiles for measles depending on the time post infection (taken from \cite{misin2020measles}); (b) For different strains of COVID-19 (e.g., WT, Delta, Omicron), the dynamic change in TCID$_{50}$/mL count depending on the time post infection (taken form \cite{hui2022sars}).}
\label{viral_load_1}
\end{center}
\end{figure}

To accurately understand the progression of an epidemic, it is crucial to consider the dynamic interplay between the pathogen and the individual's immune system following infection. The infectivity of an individual is directly influenced by their immunological response, which can vary over time post infection. For a more precise understanding, the disease transmission rate should be modeled as a function of time post infection, incorporating various immunological indicators such as TCID$_{50}$, IgG, IgM, viral load, and others. In the case of measles, the trends in IgG and IgM profiles depending on the time post infection are illustrated in \cite{misin2020measles} (see Figure-\ref{viral_load_1}). Similarly, in \cite{hui2022sars}, for different strains of COVID-19, including the wild type (WT), Delta, and Omicron variants, experimental data demonstrate the dynamic changes in TCID$_{50}$/mL count depending on the time post infection (see Figure \ref{viral_load_1}). Future research should focus on constructing transmission rates that depend on time post infection by utilizing such immunological information. Then, a more accurate expression for the basic reproduction number can be formulated as:
\begin{equation}\label{r0_distributed_beta}
\mathcal{R}_0^1 = \frac{S_0}{N} \int_{\tau}^{\infty} \beta(\eta) \eta \;(r(\eta) + d(\eta)) \, d\eta.
\end{equation}
This refined calculation of $\mathcal{R}_0$ incorporates the time-since-infection dependent transmission rate, $\beta(\eta)$, and integrates over the period $\eta$ where the infection is active, weighted by the combined recovery and death rates, $r(\eta)$ and $d(\eta)$. This approach provides a more precise determination of the basic reproduction number, which is crucial for accurately calculating the herd immunity threshold (HIT) necessary to eradicate vaccine-preventable diseases (VPDs).

Although we have estimated the time-since-infection dependent recovery and death rates, $r(\eta)$ and $d(\eta)$, our model still requires the time-since-infection dependent transmission rate and the exposed period to be fully comprehensive. The time-since-infection dependent transmission rate, $\beta(\eta)$, can be derived from immunological data on pathogen growth within an infected individual. Understanding how the pathogen load changes over time and influences transmission is essential for refining our model. These aspects, including obtaining and integrating such detailed immunological data, are areas left for future research. Exploring these elements will enhance the accuracy of our model and improve our ability to predict and control the spread of VPDs.

In this work, using various mathematical and statistical toolkits, we propose a model that offers enhanced accuracy in describing epidemic progression and provides clear insights into recovery and death distributions. In the context of modeling, we would like to mention that such modeling can be used for well-known statistical methodologies like hypothesis testing or supervised learning. For example, different strains of a particular VPD may have different features, leading to a reasonable statistical hypothesis problem in checking whether different strains have the same effect or not. From a supervised learning point of view, for a new patient affected with an unknown strain of a particular VPD, one may be interested in checking whether the unknown strain is strain A or strain B if the training dataset is available with the features of strains A and B. All these works can be of interest to future research.

\vspace{2em}

\section*{Disclaimer}
The work/opinion is based on research findings by the authors and not the opinion of the government.

\vspace{2em}

\section*{Acknowledgement}
Funding for this study was
provided by the Bill \& Melinda Gates Foundation (NV-044445).

\vspace{2em}

\section*{Data and code availability}
The sources of data used in this work are mentioned in the paper. The related codes are available in the github link:  \url{https://github.com/nsamiran/Recovery_death_distribution_estimation.git}.

\vspace{2em}



\begin{thebibliography}{10}

\bibitem{MR3197716}
Tiberiu Harko, Francisco S.~N. Lobo, and M.~K. Mak.
\newblock Exact analytical solutions of the susceptible-infected-recovered ({SIR}) epidemic model and of the {SIR} model with equal death and birth rates.
\newblock {\em Applied Mathematics and Computation}, 236:184--194, 2014.

\bibitem{bmb}
Samiran Ghosh, Vitaly Volpert, and Malay Banerjee.
\newblock An epidemic model with time-distributed recovery and death rates.
\newblock {\em Bulletin of Mathematical Biology}, 84(8):78, 2022.

\bibitem{clancy1999outcomes}
Damian Clancy.
\newblock Outcomes of epidemic models with general infection and removal rate functions at certain stopping times.
\newblock {\em Journal of Applied Probability}, 36(3):799--813, 1999.

\bibitem{lloyd2001destabilization}
Alun~L Lloyd.
\newblock Destabilization of epidemic models with the inclusion of realistic distributions of infectious periods.
\newblock {\em Proceedings of the Royal Society of London. Series B: Biological Sciences}, 268(1470):985--993, 2001.

\bibitem{dessie2022assessment}
Anteneh~Mengist Dessie, Sefineh~Fenta Feleke, Denekew~Tenaw Anley, Rahel~Mulatie Anteneh, and Zelalem~Animut Demissie.
\newblock Assessment of factors affecting time to recovery from covid-19: A retrospective study in ethiopia.
\newblock {\em Advances in Public Health}, 2022(1):7182517, 2022.

\bibitem{Verity}
Robert Verity, Lucy~C Okell, Ilaria Dorigatti, Peter Winskill, Charles Whittaker, Natsuko Imai, Gina Cuomo-Dannenburg, Hayley Thompson, Patrick~GT Walker, Han Fu, et~al.
\newblock Estimates of the severity of coronavirus disease 2019: a model-based analysis.
\newblock {\em The Lancet Infectious Diseases}, 20(6):669--677, 2020.

\bibitem{terefe2018modeling}
AN~Terefe and LA~Gebrewold.
\newblock Modeling time to recovery of adult tuberculosis (tb) patients in mizan-tepi university teaching hospital, south-west ethiopia.
\newblock {\em Mycobacterial Diseases}, 8(258):2161--1068, 2018.

\bibitem{wesley1978immunological}
Anne Wesley, HM~Coovadia, and Linda Henderson.
\newblock Immunological recovery after measles.
\newblock {\em Clinical and Experimental Immunology}, 32(3):540, 1978.

\bibitem{typhoid_info}
\url{https://my.clevelandclinic.org/health/diseases/17730-typhoid-fever}.

\bibitem{jmb}
Samiran Ghosh, Vitaly Volpert, and Malay Banerjee.
\newblock An age-dependent immuno-epidemiological model with distributed recovery and death rates.
\newblock {\em Journal of Mathematical Biology}, 86(2):21, 2023.

\bibitem{worldometer}
\url{https://www.worldometers.info/coronavirus/}.

\bibitem{MR1383587}
J.~Fan and I.~Gijbels.
\newblock {\em Local polynomial modelling and its applications}, volume~66 of {\em Monographs on Statistics and Applied Probability}.
\newblock Chapman \& Hall, London, 1996.

\bibitem{MR0172400}
\`E.~A. Nadaraja.
\newblock On non-parametric estimates of density functions and regression.
\newblock {\em Akademiya Nauk SSSR. Teoriya Veroyatnosteĭ i ee Primeneniya}, 10:199--203, 1965.

\bibitem{MR0185765}
Geoffrey~S. Watson.
\newblock Smooth regression analysis.
\newblock {\em Sankhy\=a{} Series A}, 26:359--372, 1964.

\bibitem{MR4511147}
Subhra~Sankar Dhar, Prashant Jha, and Prabrisha Rakshit.
\newblock The trimmed mean in non-parametric regression function estimation.
\newblock {\em Theory of Probability and Mathematical Statistics}, (107):133--158, 2022.

\bibitem{MR1920390}
L\'aszl\'o{} Gy\"orfi, Michael Kohler, Adam Krzy\.zak, and Harro Walk.
\newblock {\em A distribution-free theory of nonparametric regression}.
\newblock Springer Series in Statistics. Springer-Verlag, New York, 2002.

\bibitem{MR0848134}
B.~W. Silverman.
\newblock {\em Density estimation for statistics and data analysis}.
\newblock Monographs on Statistics and Applied Probability. Chapman \& Hall, London, 1986.

\bibitem{bailey1954statistical}
Norman~TJ Bailey.
\newblock A statistical method of estimating the periods of incubation and infection of an infectious disease.
\newblock {\em Nature}, 174(4420):139--140, 1954.

\bibitem{chowell_book}
Gerardo Chowell, James~M Hyman, Lu{\'\i}s~MA Bettencourt, Carlos Castillo-Chavez, and H~Nishiura.
\newblock {\em Mathematical and statistical estimation approaches in epidemiology}.
\newblock Springer, 2009.

\bibitem{survey_data}
\url{https://github.com/mrc-ide/COVID19\_CFR\_submission}.

\bibitem{chan2021sars}
Michael~CW Chan, Kenrie~PY Hui, John Ho, Man-chun Cheung, Ka-chun Ng, Rachel Ching, Ka-ling Lai, Tonia Kam, Haogao Gu, Ko-Yung Sit, et~al.
\newblock Sars-cov-2 omicron variant replication in human respiratory tract ex vivo.
\newblock \url{https://doi.org/10.21203/rs.3.rs-1189219/v1}, 2021.

\bibitem{strain_specific_data}
\url{https://ourworldindata.org/grapher/covid-cases-omicron?time=2022-01-24&country=GBR~FRA~BEL~DEU~ITA~ESP~USA~ZAF~BWA~AUS}.

\bibitem{measles_data_UK}
\url{https://github.com/NThakkar-IDM/uk_measles_surveillance/tree/main/_data}.

\bibitem{kanungo2008epidemiology}
Suman Kanungo, Shanta Dutta, and Dipika Sur.
\newblock Epidemiology of typhoid and paratyphoid fever in india.
\newblock {\em The Journal of Infection in Developing Countries}, 2(06):454--460, 2008.

\bibitem{misin2020measles}
Andrea Misin, Roberta~Maria Antonello, Stefano Di~Bella, Giuseppina Campisciano, Nunzia Zanotta, Daniele~Roberto Giacobbe, Manola Comar, and Roberto Luzzati.
\newblock Measles: an overview of a re-emerging disease in children and immunocompromised patients.
\newblock {\em Microorganisms}, 8(2):276, 2020.

\bibitem{hui2022sars}
Kenrie~PY Hui, John~CW Ho, Man-chun Cheung, Ka-chun Ng, Rachel~HH Ching, Ka-ling Lai, Tonia~Tong Kam, Haogao Gu, Ko-Yung Sit, Michael~KY Hsin, et~al.
\newblock Sars-cov-2 omicron variant replication in human bronchus and lung ex vivo.
\newblock {\em Nature}, 603(7902):715--720, 2022.

\end{thebibliography}

\newpage

\section*{Supplementary materials-I}
\noindent {\bf Epidemiological data sources:}\; 

\vspace{0.1in}

\begin{figure}[ht!]
\begin{center}
\mbox{
\subfigure[]{\includegraphics[scale=0.37]{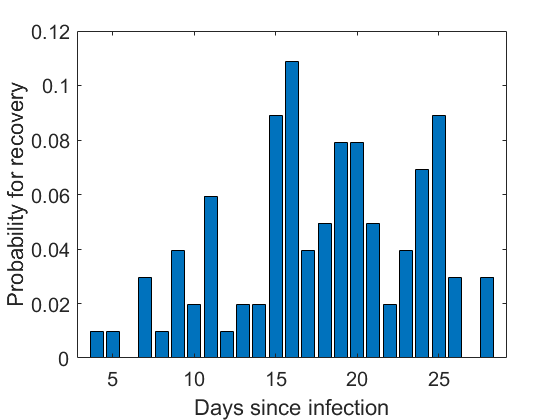}}
\subfigure[]{\includegraphics[scale=0.37]{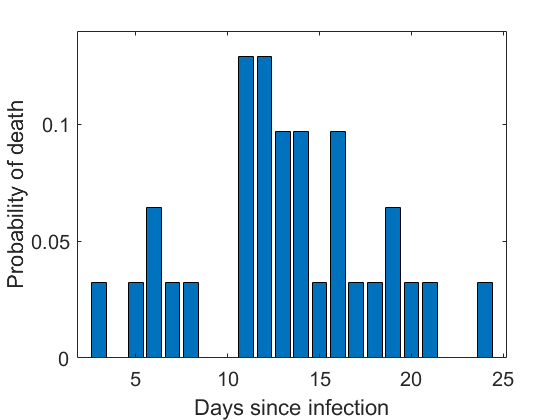}}}
\caption{(a) Probability for onset-to-recovery, (b) Probability for onset-to-death.}
\label{bar_1}
\end{center}
\end{figure}

\noindent In this study, we utilize the time series data encompassing daily number of new cases, daily number of recovery and death, number of active cases in the context of COVID-19 for different countries and different strains. These data sets have been sourced from \cite{worldometer}. Furthermore, in order to validate our proposed estimation technique, we have employed a survey data specific to the COVID-19 situation in Mainland China, accessible through \cite{survey_data,Verity} (see Figure.~\ref{bar_1}). Moreover, this study incorporates biweekly measles data from Nottingham and Derby regions in UK, sourced from \cite{measles_data_UK} and the data of typhoid for the region Kolkata in India \cite{kanungo2008epidemiology}.\\

\section*{Supplementary materials-II}
 \subsection{Estimation procedure for $d(\eta)$}

Let $f_d(t)$ be the gamma pdf describing the probability of death as a function of the time-since-infection for an infected individual who will die due to infection. Since, $p_0$ is the survival probability of infected individuals corresponding to a specific infection, we can write $$d(t) = (1-p_0) f_d(t),$$ 
with
$$f_d(t)=\frac{1}{b_d^{a_d} \Gamma (a_d)} t^{a_d-1} e^{-\frac{t}{b_d}}, $$
where, the shape parameter $a_d$ is assumed to be greater than 1, and $b_d>0$ is the scale parameter.

Define
\begin{equation}\label{Dncap_1}
\widehat{D}_{new}(t;a_d,b_d):=\int_0^t (1-p_0) f_d(t-\eta) \widehat{J}(\eta) d \eta,\;\;\;\;\text{for} \;\;t \in [0, T],
\end{equation}
where, $\widehat{J}(\eta)$ is given in \eqref{jhat_1}.

Now suppose that we have real data of daily number of new death $D_n$, at the time points $\overline{t}_1, \overline{t}_2, \cdots, \overline{t}_{k},$ and the data are represented by $\overline{D}_{{new}_1}, \overline{D}_{{new}_2}, \cdots, \overline{D}_{{new}_k}.$ Again as in the previous subsection, in principle, the two sets 
$$\big\{ \overline{t}_1, \overline{t}_2, \cdots, \overline{t}_{k} \big\} \;\; \text{and} \;\; \big\{ t_1, t_2, \cdots, t_n \big\}$$
may not be equal, but all lie in the time interval $[0, T]$.

Suppose that the shape parameter $a_d$ and scale parameters $b_d$ varies in the feasible region $[0, C]$ and $[0, D]$ respectively, for some positive constants $C, D$. Define the function
$$E_d:[0, C] \times [0, D] \to \mathbb{R}_{\geq0}$$
by
\begin{equation}\label{Ed_1}
E_d(a_d,b_d)=\sum_{j=1}^{k} \Big( \widehat{D}_{new}(\overline{t}_j;a_d,b_d)-\overline{D}_{{new}_j} \Big)^2,\;\; \forall (a_d,b_d) \in [0, C] \times [0, D].
\end{equation}
 Note that $E_d$ is a continuous function on a compact set $[0, C] \times [0, D]$, and hence attains its global minimum in the domain $[0, C] \times [0, D]$. 
 
 Let $E_d$ attains its minima at $(a^*_d, b^*_d)$ and denote 
 \begin{equation}\label{eqb_optimal_d}
  d^*(t) = (1-p_0) f^*_d(t),
  \end{equation}
with
\begin{equation}\label{eqb_optimal_d1}
f^*_d(t)=\frac{1}{{(b^*_d)}^{a^{*}_d} \Gamma (a^{*}_d)} t^{a^{*}_d-1} e^{-\frac{t}{b^{*}_d}}.
\end{equation}

The algorithm- \ref{alg:dn} shows how to estimate $d(\eta)$.
\begin{algorithm}
    \SetAlgoLined
    \SetKwInOut{Input}{Input}\SetKwInOut{Output}{Output}
    \Input{Data $\{t_i\}$, $\{J_i\}$, $\{\overline{t}_j\}$, $\{\overline{D}_{{new}_j}\}$, where $i=0, 1, \cdots, n$, $j=0, 1, \cdots, k$ }
    \Output{$(a^*_d, b^*_d)$ and $d^*(t)$}

    \nl Define time interval $[0, T]$.

    \nl Calculate the Nadaraya-Watson Estimator $\widehat{J}(\xi)$ using the formula \eqref{jhat_1}.

    \nl Define the survival probability $p_0$.

    \nl Define the feasible range of the shape ($a_d$) and scale ($b_d$) parameters, i.e., $(a_d,b_d) \in [0,C]\times [0,D]$.

    \nl Define $T_d^l$ and $T_d^u$.

    \nl Create a mesh grid for the shape ($a_d$) and scale ($b_d$) parameters in the domain $[0, C]\times [0,D]$.

    \nl \For {each combination of $(a_d,b_d)$ with $T_d^l \leq (a_d-1)b_d \leq T_d^u$}{

        \nl Calculate $\widehat{D}_{new}(t;a_d,b_d)$ as defined in \eqref{Dncap_1}.

        \nl Calculate the error term $E_d(a_d,b_d)$ as defined in \eqref{Ed_1}.
        
    }

    \nl Calculate the minimum of $E_d$ in the whole mesh and denote the corresponding mesh grid by $(a_d^*,b_d^*)$.

    \nl Compute $f_d^*(t)$ and $d^*(t)$ using the formula \eqref{eqb_optimal_d1} and \eqref{eqb_optimal_d}.
\caption{Estimation of $d(\eta)$}
\label{alg:dn}
\end{algorithm}

\newpage

\section*{Supplementary materials-III}

\begin{figure}[ht!]
\begin{center}
\mbox{
\subfigure[]{\includegraphics[scale=0.35]{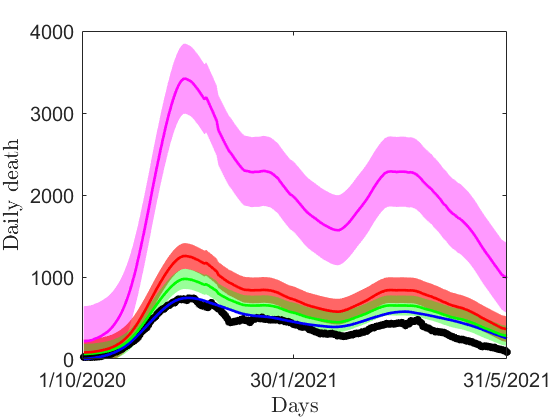}}
\subfigure[]{\includegraphics[scale=0.35]{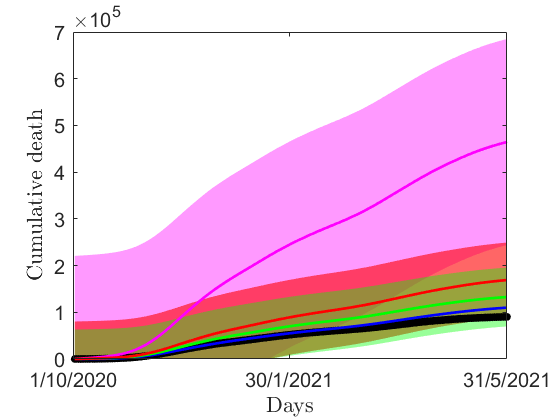}}}
\mbox{
\subfigure[]{\includegraphics[scale=0.35]{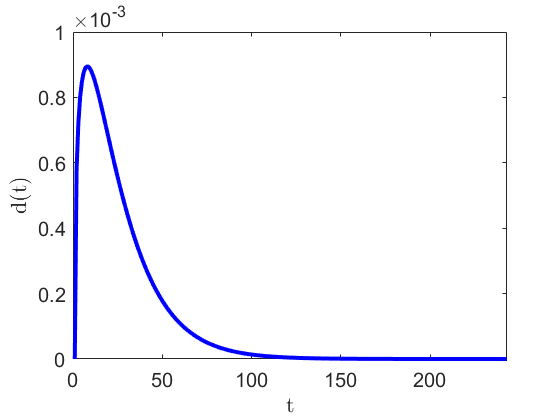}}}
\caption{The black dots correspond to real data for COVID-19 in Italy during 1/10/2020 to 31/5/2021, when the Alpha strain was dominant. The blue curves in panel (a), (b) and (c) correspond to $\widehat{D}_{new}(t;a^*_d,b^*_d)$ in formula \eqref{Dncap_1}, the cumulative $\widehat{D}_{new}(t;a^*_d,b^*_d)$, and the $d^*(t)$ respectively, with the optimal situation $(a^*_d,b^*_d)=(1.4,17.55)$. The daily death from the classical SIR-type model is given by $d_0I(t)$, where $I(t)$ is the active cases at time $t$, and $d_0=(1-p_0)/ mean$ (green curve); $d_0=(1-p_0)/ median$ (red curve); $d_0=(1-p_0)/ mode$ (magenta curve). The 3-sigma range for the green, red and magenta curves are shown by the shaded region with corresponding transparent colors.
}
\label{italy_estimate_1_d}
\end{center}
\end{figure}

\begin{figure}[ht!]
\begin{center}
\mbox{
\subfigure[]{\includegraphics[scale=0.35]{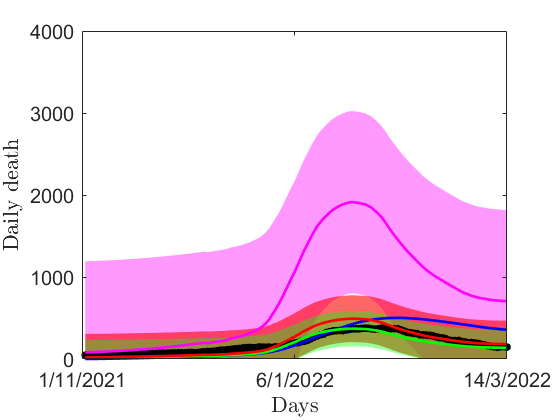}}
\subfigure[]{\includegraphics[scale=0.35]{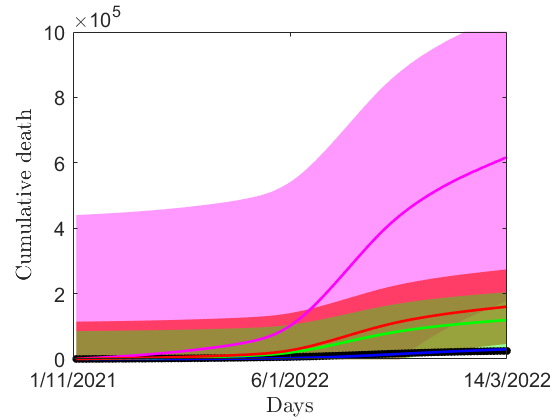}}}
\mbox{
\subfigure[]{\includegraphics[scale=0.35]{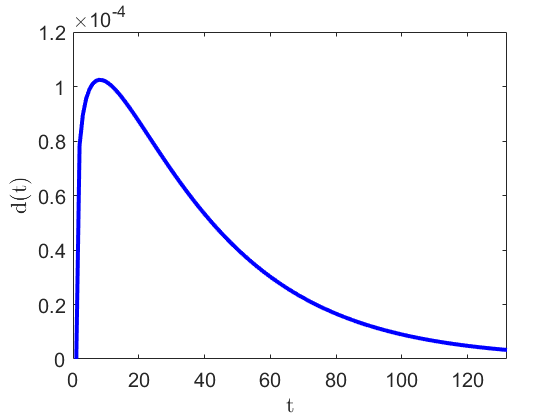}}}
\caption{The black dots correspond to real data for COVID-19 in Italy during 1/11/2021 to 14/3/2022, when the Delta strain was dominant. The blue curves in panel (a), (b) and (c) correspond to $\widehat{D}_{new}(t;a^*_d,b^*_d)$ in formula \eqref{Dncap_1}, the cumulative $\widehat{D}_{new}(t;a^*_d,b^*_d)$, and the $d^*(t)$ respectively, with the optimal situation $(a^*_d,b^*_d)=(1.24,30)$. The daily death from the classical SIR-type model is given by $d_0I(t)$, where $I(t)$ is the active cases at time $t$, and $d_0=(1-p_0)/ mean$ (green curve); $d_0=(1-p_0)/ median$ (red curve); $d_0=(1-p_0)/ mode$ (magenta curve). The 3-sigma range for the green, red and magenta curves are shown by the shaded region with corresponding transparent colors.}
\label{italy_estimate_2_d}
\end{center}
\end{figure}

\begin{figure}[ht!]
\begin{center}
\mbox{
\subfigure[]{\includegraphics[scale=0.35]{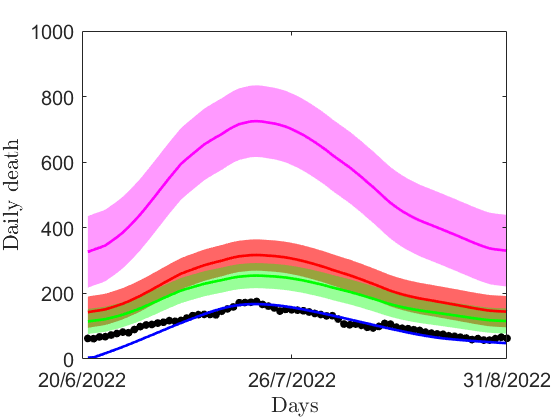}}
\subfigure[]{\includegraphics[scale=0.35]{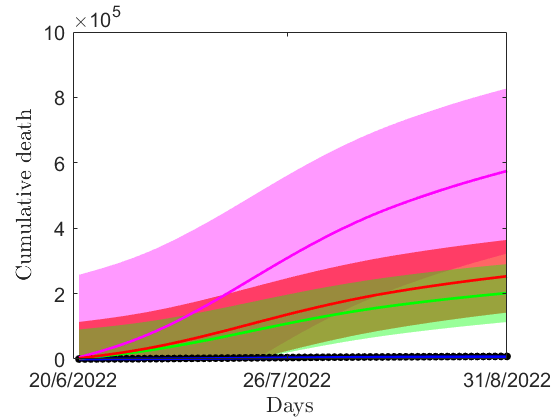}}}
\mbox{
\subfigure[]{\includegraphics[scale=0.35]{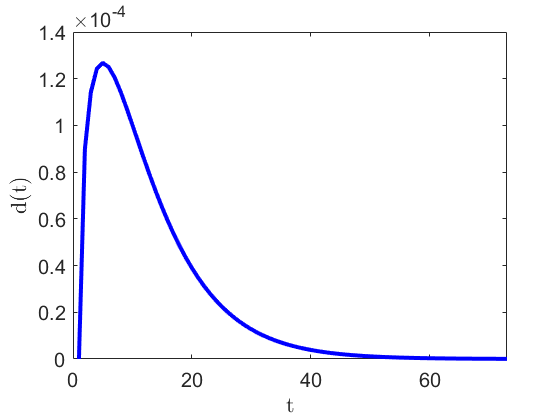}}}
\caption{The black dots correspond to real data for COVID-19 in Italy during 20/6/2022 to 31/8/2022, when the Omicron strain was dominant. The blue curves in panel (a), (b) and (c) correspond to $\widehat{D}_{new}(t;a^*_d,b^*_d)$ in formula \eqref{Dncap_1}, the cumulative $\widehat{D}_{new}(t;a^*_d,b^*_d)$, and the $d^*(t)$ respectively, with the optimal situation $(a^*_d,b^*_d)=(1.54,7.42)$. The daily death from the classical SIR-type model is given by $d_0I(t)$, where $I(t)$ is the active cases at time $t$, and $d_0=(1-p_0)/ mean$ (green curve); $d_0=(1-p_0)/ median$ (red curve); $d_0=(1-p_0)/ mode$ (magenta curve). The 3-sigma range for the green, red and magenta curves are shown by the shaded region with corresponding transparent colors.}
\label{italy_estimate_3_d}
\end{center}
\end{figure}



\end{document}